\documentclass[11pt,twoside]{article}
\usepackage{asp2010}

\begin{document}

\title{What’s in a Fermi Bubble: a quasar episode in the Galactic centre}
\author{Kastytis Zubovas$^{1}$, Sergei Nayakshin$^1$, Andrew R. King$^1$
  \affil{$^1$ Dept. of Physics \& Astronomy, University of Leicester,
    Leicester, LE1 7RH, UK; mailto:kastytis.zubovas@le.ac.uk}}

\begin{abstract}

{\it Fermi} bubbles, the recently observed giant ($\sim10$~kpc high) gamma-ray
emitting lobes on either side of our Galaxy \citep{Su2010ApJ}, appear
morphologically connected to the Galactic center, and thus offer a chance to
test several models of supermassive black hole (SMBH) evolution, feedback and
relation with their host galaxies.  We use a physical feedback model
\citep{King2003ApJ, King2010MNRASa} and novel numerical techniques
\citep{Nayakshin2009MNRASb} to simulate a short burst of activity in
Sgr~A$^*$, the central SMBH of the Milky Way, $\sim6$~Myr ago, temporally
coincident with a star formation event in the central parsec. We are able to
reproduce the bubble morphology and energetics both analytically
\citep{Zubovas2011MNRAS} and numerically (Zubovas \& Nayakshin, in
prep). These results provide strong support to the model, which was also used
to simulate more extreme environments \citep{Nayakshin2010MNRAS}.

\end{abstract}

The AGN radiation radiation pressure drives a wind with a momentum flux
$\dot{M}_{\rm out}v \simeq L_{\rm Edd}/c$ with $v \simeq \eta c \simeq 0.1c$,
where $\eta \simeq 0.1$ is the radiative efficiency \citep{King2003ApJ}. This
wind shocks against the surrounding gas (perhaps producing $\gamma$ rays) and
pushes it away, forming an outflow. In the Milky Way, the wind shock cannot
cool outside $R_{\rm cool} \sim 10$~pc and hence transfers most of the kinetic
energy rate ($\sim 0.05L_{\rm Edd}$) to the ambient gas (this is an
energy-driven flow). Such an outflow, while driven, moves with a constant
velocity $v_e \sim 1000$~km~s$^{-1}$ \citep{King2011MNRAS}. Once the quasar
switches off, the shell coasts for an order of magnitude longer than the
driving phase $t_{\rm q}$, easily reaching radii of tens of kpc.

The outflow morphology can become non-spherical due to anisotropic matter
distribution in the Galaxy, such as the dense gas in the Central Molecular
Zone (CMZ) which is too heavy for even an energy-driven outflow to lift. This
qualitatively explains the morphology of the {\it Fermi} bubbles. We use their
observed and inferred properties to constrain the gas fraction (ratio of gas
density to background potential density) in the Galaxy halo and the Sgr~A$^*$
outburst duration \citep[see][for details]{Zubovas2011MNRAS}:
\begin{equation}
f_{\rm g} \lesssim 7\cdot 10^{-3}l; \;\; t_{\rm q} > 2.5 \cdot 10^5 \; {\rm yr}.
\end{equation}

\begin{figure}
  \plottwo{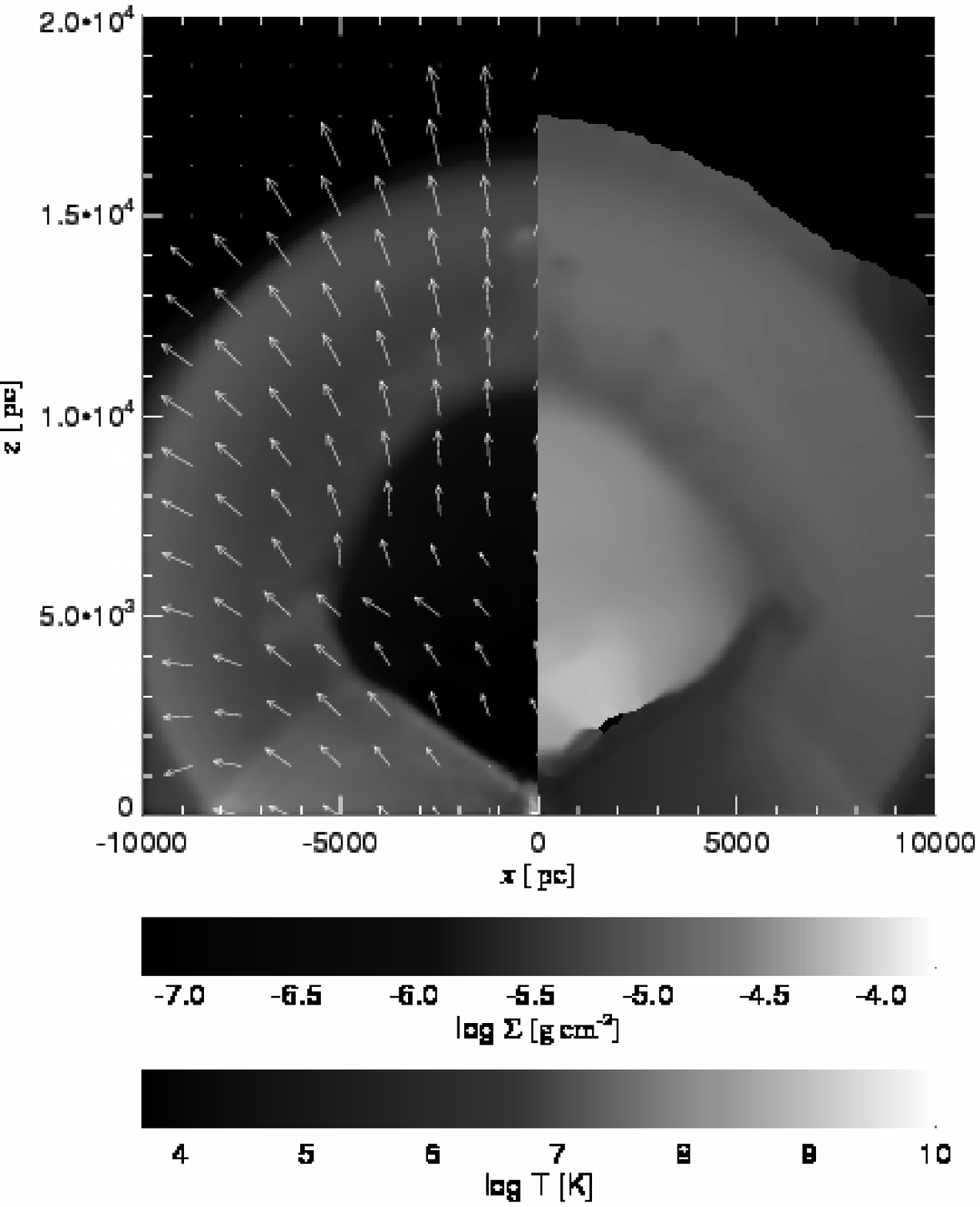}{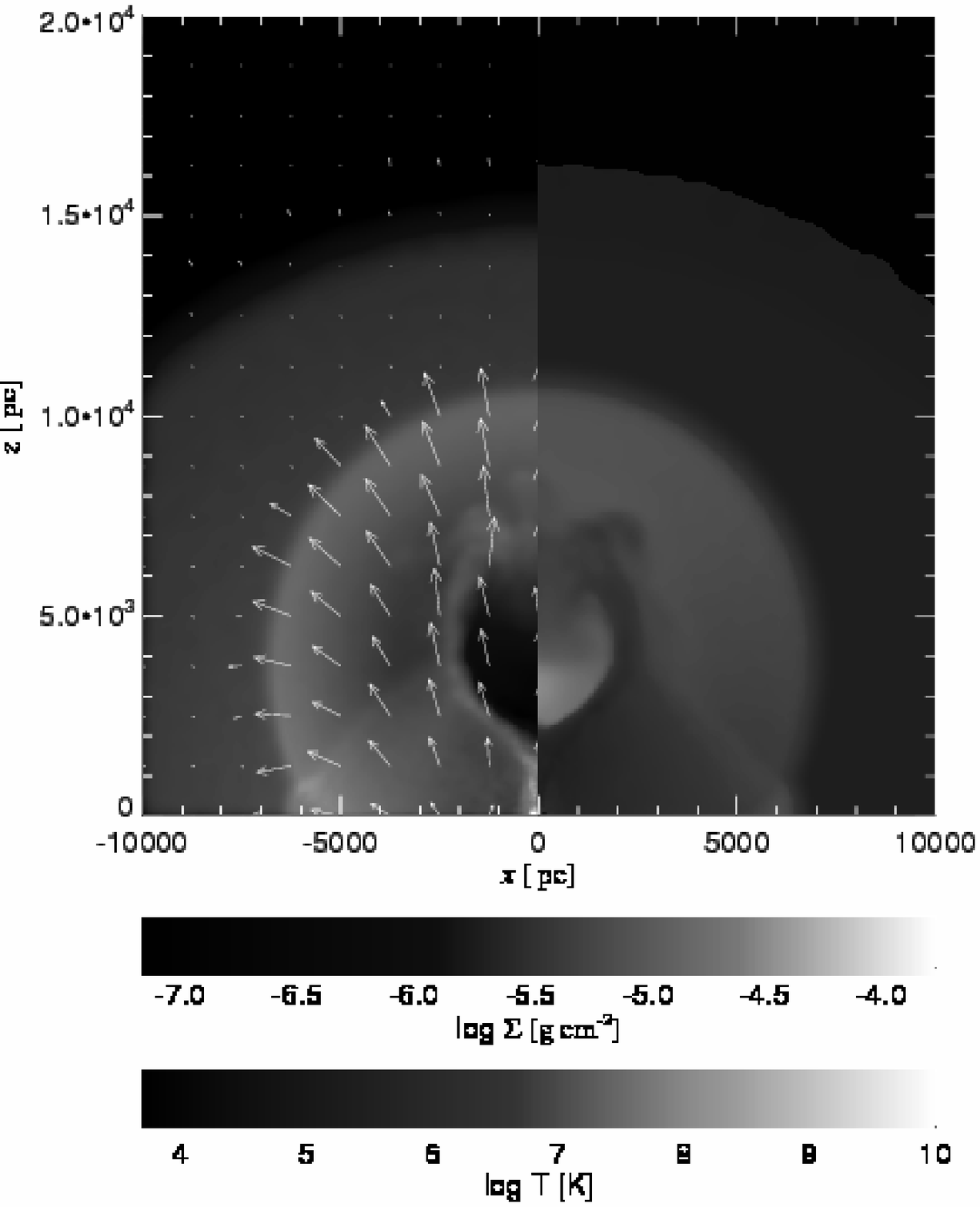}
  \caption{{\it Left}: Gas surface density (left) and temperature (right) at
    $t = 6$~Myr. The outflow is collimated and forms teardrop-shaped cavities
    with similar morphology to that of the observed {\it Fermi} bubbles. The
    simulation has $t_{\rm q} = 1$~Myr, $f_{-3} = 1$. {\it Right}: Same as
    left, but $t_{\rm q} = 0.3$~Myr. The cavity morphology is obviously
    inconsistent with observations. A change in $f_{\rm g}$ produces
    inconsistent results as well.}
  \label{fig1}
\end{figure}

We test the model numerically, using GADGET with a 'virtual particle' method
of implementing wind feedback \citep{Nayakshin2009MNRASb}. We embed the SMBH
(which produces feedback for a time $t_{\rm q}$) and CMZ into a spherically
symmetric isothermal halo with $\sigma = 100$~km/s and a constant $f_{\rm
  g}$. We vary the free parameters $t_{\rm q}$ and $f_{\rm g}$.

Figure \ref{fig1}, left, shows that our model, with $t_{\rm q} = 1$~Myr and
$f_{\rm g} = 10^{-3}$, can reproduce the morphology and size of the observed
{\it Fermi} bubbles. The CMZ is perturbed but not dispersed by the wind and
collimates the outflow into two cavities. The total energy content inside the
cavities is a small fraction of the input and also agrees with observational
constraints. Figure \ref{fig1}, right, shows a simulation with $t_{\rm q} =
0.3$~Myr, which produces bubbles clearly inconsistent with observations. We
can thus put tight constraints on both parameters.  We also require the CMZ
mass to be $\simeq10^8 M_{\odot}$, but its aspect ratio is not important. A
physical heating-cooling prescription \citep{Sazonov2005MNRAS} does not change
the results significantly either.  Therefore our findings are quite robust
with regard to the uncertainties involved in the initial conditions.

We have shown that our physically motivated SMBH wind feedback model can
explain the {\it Fermi} bubbles. In addition, the same model works for quasars
as their SMBHs establish the $M-\sigma$ relation and clear the host galaxies
of gas \citep[e.g.][]{Nayakshin2010MNRAS}, suggesting there is no fundamental
difference between the processes that were responsible for forming the
galaxies at $z \gtrsim 2$ and the processes what is happening in local, mostly
dormant, galactic nuclei.

\acknowledgments This research used the ALICE High Performance Computing
Facility at the University of Leicester.  
KZ is supported by an STFC studentship.

\end{document}